\documentclass[12pt]{article}
\usepackage{epsfig}
\def\beq{\begin{equation}}
\def\eeq{\end{equation}}
\def\ga{\mathrel{\raise.3ex\hbox{$>$\kern-.75em\lower1ex\hbox{$\sim$}}}}
\def\la{\mathrel{\raise.3ex\hbox{$<$\kern-.75em\lower1ex\hbox{$\sim$}}}}

\def\bc{\begin{center}}
\def\ec{\end{center}}
\def\bea{\begin{eqnarray}}
\def\eea{\end{eqnarray}}

\def\gappeq{\mathrel{\rlap {\raise.5ex\hbox{$>$}} {\lower.5ex\hbox{$\sim$}}}}
\def\lappeq{\mathrel{\rlap{\raise.5ex\hbox{$<$}} {\lower.5ex\hbox{$\sim$}}}}

\catcode`@=11
\def\marginnote#1{}
\newcount\hour
\newcount\minute
\newtoks\amorpm
\hour=\time\divide\hour by60
\minute=\time{\multiply\hour by60 \global\advance\minute by-\hour}
\edef\standardtime{{\ifnum\hour<12 \global\amorpm={am}%
        \else\global\amorpm={pm}\advance\hour by-12 \fi
        \ifnum\hour=0 \hour=12 \fi
        \number\hour:\ifnum\minute<10 0\fi\number\minute\the\amorpm}}
\edef\militarytime{\number\hour:\ifnum\minute<10 0\fi\number\minute}
\def\draftlabel#1{{\@bsphack\if@filesw {\let\thepage\relax
   \xdef\@gtempa{\write\@auxout{\string
      \newlabel{#1}{{\@currentlabel}{\thepage}}}}}\@gtempa
   \if@nobreak \ifvmode\nobreak\fi\fi\fi\@esphack}
        \gdef\@eqnlabel{#1}}
\def\@eqnlabel{}
\def\@vacuum{}
\def\draftmarginnote#1{\marginpar{\raggedright\scriptsize\tt#1}}
\def\draft{\oddsidemargin 0.0truein
        \def\@oddfoot{\sl preliminary draft \hfil
        \rm\thepage\hfil\sl\today\quad\militarytime}
        \let\@evenfoot\@oddfoot \overfullrule 3pt
        \let\label=\draftlabel
        \let\marginnote=\draftmarginnote
   \def\@eqnnum{(\theequation)\rlap{\kern\marginparsep\tt\@eqnlabel}%
\global\let\@eqnlabel\@vacuum}  }
\catcode`@=12
\begin{document}
\begin{titlepage}
\vspace*{-1cm}
\rightline{CERN-TH/2002-135}
\rightline{UMN--TH--2102/02}
\rightline{TPI--MINN--02/17}
\rightline{hep-ph/0206163} 
\vskip 2.0cm
\begin{center}
{\Large\bf Astrophysical and Cosmological Constraints on Neutrino masses
\footnote{to appear in ``Neutrino Mass'', Springer Tracts in Modern Physics, ed. by G. Altarelli and
K. Winter.}}
%\\
%\vskip .1cm
%Brane Action and Symmetry Breaking}
\end{center}
\vskip 1.5  cm
\begin{center}
{\large Kimmo Kainulainen}~\footnote{e-mail address: kimmo.kainulainen@cern.ch;
Permanent address: University of
Jyv\"askyl\"a, Finland.}
\\
\vskip .1cm
Theory Division, CERN, CH-1211, Geneva, Switzerland  and \\ NORDITA,
Blegdamsvej  17, DK-2100, Copenhagen \O , Denmark

\vskip .5cm
{\large Keith A. Olive}~\footnote{e-mail address: olive@umn.edu}
\\
\vskip .1cm
Theoretical Physics Institute, School of Physics and
Astronomy,\\  University of Minnesota, Minneapolis, MN 55455, USA
\end{center}
\vskip 1.5cm
\begin{abstract}
\noindent
We review some astrophysical and cosmological properties and implications of neutrino
masses and mixing angles. These include: constraints based on the relic density of
neutrinos, limits on their masses and lifetimes, BBN limits on mass parameters,
neutrinos and supernovae, and neutrinos and high energy cosmic rays.
\end{abstract}
\end{titlepage}
\setcounter{footnote}{0}
\vskip2truecm
\section{Introduction}

The role of neutrinos in cosmology and astrophysics can not be
understated \cite{dolgovreview}.  They play a critical role in the physics
of the early Universe, at temperatures scales of order 1 MeV, and strongly
determine the abundances of the light elements produced in big bang
nucleosynthesis. They almost certainly play a key role in supernova
explosions, and if they have mass, could easily contribute to the 
overall mass density of the Universe. At the present time,
the only indicators of neutrino masses are from astrophysical sources,
the inferred neutrino oscillations of neutrinos produced in the Sun, and
those produced in cosmic-ray collisions in the atmosphere. 
Indeed, their elusive character has proven that a great deal of
information on neutrino properties can be gained by studying their
behavior in astrophsyical and cosmological environments.
Here, we will try to elucidate some of these constraints on neutrino
masses.

In our discussion below, we will assume that the early Universe
is well described by a standard Friedmann-Lemaitre-Robertson-Walker
metric 
\beq
  ds^2 = dt^2 - R^2(t) \left [ \frac{d r ^2}{1 - k r ^2} + r^2\,
  (d \theta^2 + \sin^2\theta\, d \phi^2 ) \right ] 
\eeq
We further assume that thermal equilibrium was established at some early
epoch and that we can describe the radiation by a black body equation of
state, $p = \rho/3$ at a temperature $T$. Solutions to Einstein's equations 
allow one to determine the expansion rate of the Universe defined to be 
the Hubble parameter in terms of the energy density in radiation, the 
curvature and the cosmological constant. In the early Universe the latter 
two quantities can be neglected and we write
\beq
 H^2 \equiv
        \left (
        \frac{\dot R}{R}
        \right )^2 =
        \frac{8 \pi\; G_{N}\; \rho}{3}    
\label{H}  
\eeq
where the energy density is 
\beq
 \rho {} = \left( \sum_B g_{B} + {7 \over 8} \sum_F  g_{F} \right)
   {\pi^{ 2} \over 30}  T^{4} \equiv    {\pi^{ 2} \over 30}\, N(T)
\, T^{4} 
\label{rho}
\eeq
The present neutrino contribution to the total energy density, relative
to  the critical density (for a spatially flat Universe) is
\beq
\Omega_\nu = \frac{\rho_\nu}{\rho_c}
\label{omegaratio}
\eeq
where $\rho_c  = 1.06 \times 10^{-5} {h}^2 {\rm GeV/cm}^3$ and
$h = H/100$km/Mpc/s is the scaled Hubble parameter.
For a recent review of standard big bang cosmology, see~\cite{OP}.

\section{The Cosmological Relic Density of Stable Neutrinos}
\label{sec:1.1}

The simplicity of the standard big bang model allows one to
compute in a straightforward manner the relic density of any
stable particle if that particle was once in thermal equilibrium
with the thermal radiation bath.  At early times, neutrinos were
kept in thermal equilibrium by their weak interactions with electrons 
and positrons. Equilibrium is achieved whenever some rate $\Gamma$ is 
larger than the expansion rate of the Universe, or $\Gamma_{ i} > H$.
Recalling that the age of the Universe is determined by $H^{-1}$, this
condition is equivalent to requiring that on average, at least one
interaction has occurred over the life-time of the Universe. On 
dimensional grounds, one can estimate the thermally averaged 
low-energy weak interaction scattering cross section
\beq
 \langle \sigma v \rangle\; { \sim }~g^4 T^{ 2} /m_{W}^4
\eeq
for $T \ll m_W$. Recalling that the number density scales as 
$n \propto T^3$, we can compare the weak interaction rate 
$\Gamma \sim n \langle \sigma v \rangle$, with the expansion rate 
given by eqs. (\ref{H}) and (\ref{rho}). Neutrinos will be in 
equilibrium when $\Gamma_{\rm wk} > H$ or
\beq
  T^3 > \sqrt{8 \pi^3 N /90}\,\,\, m_{W}^4/M_{P}
\eeq
where $M_{P} = G_{N}^{-1/2} =  1.22 \times 10^{19}$ GeV is the Planck
mass.  For $N = 43/4$ (accounting for  photons, electrons, positrons 
and three neutrino flavors) we see that equilibrium is maintained at
temperatures greater than ${\cal O}(1)$ MeV (for a more accurate 
calculation see~\cite{Enqvist:gx}).

The decoupling scale of ${\cal O}(1)$ MeV has an important consequence 
on the final relic density of massive neutrinos. Neutrinos more massive 
than 1 MeV will begin to annihilate prior to decoupling, and while in 
equilibrium, their number density will become exponentially suppressed. 
Lighter neutrinos decouple as radiation on the other hand, and hence do 
not experience the suppression due to annihilation. Therefore, the 
calculations of the number density of light ($m_\nu \la 1$ MeV) and 
heavy ($m_\nu \ga 1$ MeV) neutrinos differ substantially.

\begin{figure}
\begin{center}
\includegraphics[height=6.0cm]{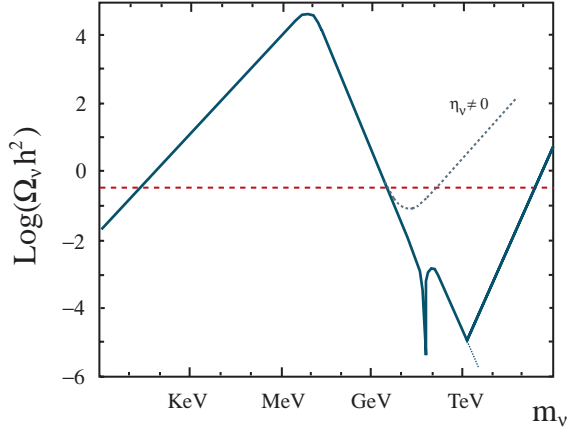}
\caption{Summary plot of the relic density of Dirac neutrinos 
(solid) including a possible neutrino asymmetry of $\eta_\nu =
5\times 10^{-11}$ (dotted).}
\label{fig:1}      
\end{center}
\end{figure}

The number of density of light neutrinos with $m_\nu \la 1$ MeV can be 
expressed at late times as
\beq
  \rho_\nu  = m_\nu Y_\nu n_\gamma  
\label{rhonus}
\eeq
where $Y_\nu = n_\nu/n_\gamma$ is the density of $\nu$'s relative to 
the density of photons, which today is 411 photons per cm$^3$. It is 
easy to show that in an adiabatically expanding universe $Y_\nu = 
3/11$. This suppression is a result of the $e^+ e^-$ annihilation 
which occurs after neutrino decoupling and heats the photon bath 
relative to the neutrinos.
In order to obtain an age of the Universe, $t > 12$ Gyr, one 
requires that the matter component is constrained by
\beq
  \Omega h^2 \le 0.3.
\label{omegabound}
\eeq  
From this one finds the strong constraint (upper bound) on Majorana
neutrino masses: \cite{cows}
\beq
m_{\rm tot} =   \sum_\nu  m_\nu   \la 28 {\rm eV}.
\label{ml1}
\eeq
where the sum runs over neutrino mass eigenstates. The limit for Dirac
neutrinos depends on the interactions of the right-handed states (see
discussion below).  As one can see, even very small  neutrino masses of
order 1 eV, may contribute substantially to the  overall relic density.
The limit (\ref{ml1}) and the corresponding  initial rise in
$\Omega_\nu h^2$ as a function of $m_\nu$ is  displayed in the
Figure~\ref{fig:1} (the low mass end with 
$m_\nu \la 1$ MeV).

\begin{figure}
\begin{center}
\includegraphics[height=5cm]{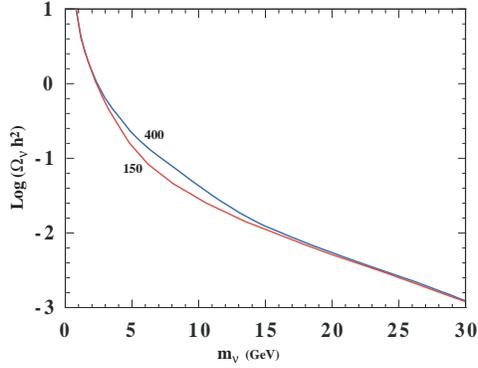}
\caption{The relic density of heavy Dirac neutrinos due to 
annihilations \protect\cite{wso}. The curves are labeled by
the assumed quark-hadron phase transition temperature in MeV.}
\label{fig:2}       
\end{center}
\end{figure}

The calculation of the relic density for neutrinos more massive than
$\sim 1$ MeV, is substantially more involved. The relic density is now
determined by the freeze-out of neutrino annihilations which occur at
$T \la m_\nu$, after annihilations have begun to seriously reduce their
number density
\cite{lw}. The annihilation rate is given by
\beq
\Gamma_{ann} = \langle \sigma v \rangle_{ann} n_\nu \sim
\frac{m_\nu^2}{m_Z^4} (m_\nu T)^{3/2} e^{-m_\nu/T}
\label{annrate}
\eeq
where we have assumed, for example, that the annihilation cross section
is dominated by $\nu {\bar \nu} \rightarrow f {\bar f}$ via $Z$-boson
exchange\footnote{While this is approximately true for Dirac neutrinos,
the annihilation cross section of Majorana neutrinos is $p$-wave
suppressed and is proportional of the final state fermion masses 
rather than $m_\nu$.} 
and $\langle \sigma v \rangle_{ann} \sim m_\nu^2/m_Z^4$. When the 
annihilation rate becomes slower than the expansion rate of the Universe
the annihilations freeze out and the relative abundance of neutrinos
becomes fixed. 
Roughly, $Y_\nu \sim (m \langle \sigma v \rangle_{ann} )^{-1}$
and hence $\Omega_\nu h^2 \sim {\langle \sigma v \rangle_{ann}}^{-1}$,
so that parametrically $\Omega_\nu h^2  \sim 1/{m_\nu^2}$. As a result,
the constraint (\ref{omegabound}) now leads to a {\em lower} 
bound~\cite{lw,ko,wso} on the neutrino mass, of about $m_\nu \ga 3-7$ 
GeV, depending on whether it is a Dirac or Majorana neutrino. 
This bound and the corresponding downward trend $\Omega_\nu h^2 \sim
1/m^2_\nu$ can again be seen in Figure~\ref{fig:1}. The result of a
more detailed calculation is shown in
Figure~\ref{fig:2}~\cite{wso} for the case of a Dirac neutrino. 
The two curves show the slight sensitivity on the temperature scale
associated with the quark-hadron transition. The result for a Majorana
mass neutrino is qualitatively similar. 
Indeed, any particle with roughly weak scale cross-sections will tend to
give an interesting  value of  $\Omega h^2 \sim 1$.

The deep drop in $\Omega_\nu h^2$, visible in Figure~\ref{fig:1}
at around $m_\nu = M_Z/2$, is due to a very strong annihilation 
cross section at $Z$-boson pole. For yet higher neutrino masses the 
$Z$-annihilation channel cross section drops as $\sim 1/m_\nu^2$, 
leading to a brief period of an increasing trend in $\Omega_\nu h^2$. 
However, for $m_\nu \ga m_W$ the cross section regains its parametric 
form $\langle \sigma v \rangle_{ann} \sim m_\nu^2$ due to the opening up
of a new annihilation channel to $W$-boson pairs~\cite{Enqvist:1988we}, 
and the density drops again as $\Omega_\nu h^2 \sim 1/m^2_\nu$. 
The tree level $W$-channel cross section breaks the unitarity at 
around ${\cal O}({\rm few})$ TeV~\cite{Enqvist:yz} however, and the full 
cross section must be bound by the unitarity limit~\cite{Griest:1989wd}.
This behaves again as $1/m_\nu^2$, whereby $\Omega_\nu h^2$ 
has to start increasing again, until it becomes too large again at
200-400 TeV~\cite{Griest:1989wd,Enqvist:yz} (or perhpas somewhat 
earlier as the weak interactions become strong at the unitarity
breaking scale).

\section{Neutrinos as Dark Matter}
\label{sec:1.2}

Based on the leptonic and invisible width of the $Z$ boson, 
experiments at LEP have determined that the number of neutrinos is 
$N_\nu = 2.9841 \pm 0.0083$ \cite{RPP}. Conversely, any new physics 
must fit within these brackets, and thus LEP excludes additional 
neutrinos (with standard weak interactions) with masses $m_\nu 
\la 45$ GeV.  Combined with the limits displayed in Figures~\ref{fig:1} 
and \ref{fig:2}, we see that the mass density of ordinary heavy
neutrinos is bound to be very small, $\Omega_\nu {h}^2 < 0.001$ for
masses  $m_\nu > 45$ GeV up to $m_\nu \sim {\cal O}(100)$ TeV.

A bound on neutrino masses even stonger than Eqn.~(\ref{ml1}) can be 
obtained from the recent observations of active-active mixing in both 
solar- and atmospheric neutrino experiments. The inferred evidence for 
$\nu_\mu-\nu_\tau$ and $\nu_e-\nu_{\mu,\tau}$ mixings are on the scales 
$m_\nu^2 \sim 1-10 \times 10^{-5}$ and $m_\nu^2  \sim 2-5 \times 10^{-3}$. 
When combined with the upper bound on the electon-like neutrino mass
$m_{\nu} <  2.8$ eV~\cite{Mainz}, and the LEP-limit on the number of
neutrino  species, one finds the constraint on the sum of neutrino masses:
\beq
0.05~{\rm eV} \la m_{\rm tot} \la 8.4~\rm eV.
\eeq
Conversely, the experimental and observational data then implies that 
the cosmological energy density of all light, weakly interacting 
neutrinos can be restricted to the range 
\beq
0.0005 \la \Omega_\nu h^2 \la 0.09.
\label{range}
\eeq
Interestingly there is now also a lower bound due to the fact that at 
least one of the neutrino masses has to be larger than the scale $m^2 
\sim 10^{-3}$ eV$^2$ set by the atmospheric neutrino data.
Combined with the results on relic mass density of neutrinos and the
LEP limits, the bound~(\ref{range}) implies that the ordinary weakly 
interacting neutrinos, once the standard dark matter 
candidate~\cite{ss}, can be ruled out completely as a dominant 
component of the dark matter.

This conclusion can be evaded if neutrinos are Dirac particles, and 
have a nonzero asymmetry however, since then the relic density could 
be governed by the asymmetry rather than by the annihilation cross section. 
Indeed, it is easy to see that the neutrino mass density corresponding 
to the asymmetry $\eta_\nu \equiv (n_\nu - n_{\bar \nu})/n_\gamma$ is 
given by~\cite{ho}
\beq
  \rho = m_\nu \eta_\nu n_\gamma ,
\eeq
which implies
\beq
  \Omega_\nu h^2 \simeq 0.004 \,\eta_{\nu 10}\, (m_\nu/{\rm GeV}).
\eeq
where $\eta_{\nu 10}\equiv 10^{10}\eta_\nu$.
We have shown the behaviour of the energy density of neutrinos
with an asymmetry by the dotted line in the Figure~\ref{fig:1}. At 
low $m_\nu$, the mass density is dominated by the symmetric, 
relic abundance of both neutrinos and antineutrinos which have already
frozen out. At higher values  of $m_\nu$, the annihilations suppress the
symmetric part of the relic  density until $\Omega_\nu h^2$
eventually becomes  dominated by the linearly increasing asymmetric
contribution. In the  figure, we have assumed an asymmetry of
$\eta_\nu \sim 5 \times  10^{-11}$ for neutrinos with standard weak 
interaction strength. In this case, $\Omega_\nu h^2$ begins to rise 
when $m_\nu \ga 20$ GeV.  Obviously, the bound (\ref{omegabound}) is 
saturated for $m_\nu = 75 \, {\rm GeV}/\eta_{\nu 10}$.

There are also other cosmolgical settings that give rise to interesting
mass constraints on the eV scale.  
Indeed, light neutrinos were problematic in cosmology long before 
the imporoved mass limits leading to (\ref{range}) were established, 
due to their effect on structure formation. Light 
particles which are still relativistic at the time of matter 
domination erase primordial perturbations
due to free streaming out to very large scales~\cite{free}. Given a
neutrino with mass $m_\nu$, the smallest surviving non-linear structures 
are determined by the Jean's mass
\beq
	M_J  = 3 \times 10^{18}  {M_\odot \over {m_\nu^2}(\rm eV)}.	
\label{mj}
\eeq
Thus, for eV mass neutrinos the large scale structures, including 
filaments and voids~\cite{nu1,nu2}, must form first and galaxies
whose typical mass scale is $\simeq 10^{12} M_\odot$ are expected to 
fragment out later. Particles with this property are termed hot dark
matter (HDM). It seemed that neutrinos were ruled out because they 
tend to produce too much large scale structure~\cite{nu3}, and 
galaxies formed too late~\cite{nu2,nu4}, at $z \le 1$, whereas 
quasars and galaxies are seen out to redshifts $z \ga 6$.

Subsequent to the demise of the HDM scenario, there was a brief 
revival for neutrino dark matter as part of a mixed dark matter
model, using now more conventional cold dark matter along with a small
component of hot (neutrino) dark matter. The motivation for doing
this was to  recover some of the lost power on large scales that 
is absent in CDM models \cite{chdm}. However, galaxies still form 
late in these models, and more importantly, almost all evidence now
points away from models with $\Omega_m = 1$, and strongly favor 
models with a cosmological constant ($\Lambda$CDM).

Combining the rapidly improving data on key cosmological 
parameters with the better statistics from large redshift surveys 
has made it possible to go a step forward along this path. It is 
now possible to set stringent limits on the light neutrino mass
density $\Omega_\nu h^2$, and hence on neutrino mass based on
the power spectrum of the Ly 
$\alpha$ forest~\cite{strong}, $m_{\rm tot} < 5.5$ eV, and the limit 
is even stronger if the total matter density, $\Omega_m$ is less 
than 0.5.  This limit has recently been improved by the
2dF Galaxy redshift \cite{2dF} survey by comparing the derived power
spectrum of fluctuations with structure formation models. 
 Focussing on the the 
presently favoured $\Lambda$CDM model, the neutrino mass bound becomes
$m_{\rm tot} < 1.8 $ eV for $\Omega_m < 0.5$.

Finally, right handed or sterile neutrinos may also contribute to the dark 
matter. The mass limits for neutrinos with less than full weak strength
interactions are relaxed \cite{OT}. For Dirac neutrinos, the upper
limit varies between 100 -- 200 eV depending on the strength of their
interactions.  For Majorana neutrinos, the limit is further relaxed to 200
-- 2000 eV. This relaxation is primarily due to the dilution of the number
density of super-weakly interacting neutrinos 
due to entropy production by decay and annihilation of massive states 
after their decoupling from equilibrium~\cite{oss}.
%: since these neutrinos
%decouple from the thermal background at higher temperatures than normal
%neutrinos, the annihilations of massive states ($> 1$ MeV) reduces the
%neutrino density relative to photons \cite{oss}. 
Such neutrinos make excellent warm dark matter candidates, albeit the 
viable mass range for galaxy formation is quite restricted~\cite{warm}.

\section{Neutrinos and Big Bang Nucleosynthesis}
\label{sec:1.3}

\begin{figure}
\begin{center}
\includegraphics[height=9cm]{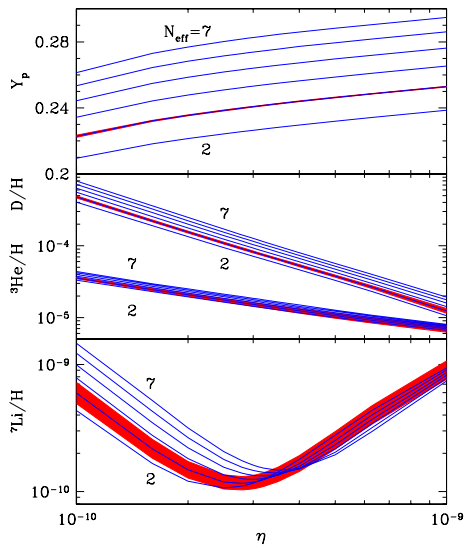}
\vskip-1.0truecm
\caption{The light element abundances as a function of the
baryon-to-photon ratio for different values of $N_\nu$
\protect\cite{cfo2}.}
\label{fig:3}      
\end{center}
\end{figure}

Big bang nucleosynthesis is the cosmological theory of the origin of 
the light element isotopes D, $^3$He, $^4$He, and $^7$Li \cite{bbn}.
The success of the theory when compared to the observational
determinations of the light elements allows one to place strong 
constraints on the physics of the early Universe at a time scale of 
1-100 seconds after the big bang.  $^4$He is the most sensitive probe 
of deviations from the standard model and its abundance is determined 
primarily by the neutron to proton ratio when nucleosynthesis begins 
at a temperature of $\sim 100$ keV (to a good approximation all 
neutrons are then bound to form $^4$He). 
The ratio $n/p$ is determined by the competition between the weak
interaction rates which interconvert neutrons and protons,

\begin{equation}
 p + e^- \leftrightarrow n + \nu_e \, , \qquad 
 n + e^+ \leftrightarrow p + \bar \nu_e \, , \qquad 
 n \leftrightarrow p + e^- + \bar \nu_e
\label{weakrates}
\end{equation}
and the expansion rate, and is largely given by the Boltzmann 
factor
\beq
  n/p \sim e^{-(m_n - m_p)/T_f}
\label{nperp}
\eeq
where $m_n-m_p$ is the neutron to proton mass difference.
As in the case of neutinos discussed above, these 
weak interactions also freeze out at a temperature of roughly 1 MeV
when
\beq
{G_F}^2 {T_f}^5 \sim \Gamma_{\rm wk}(T_f) = H(T_f) \sim \sqrt{G_N N} {T_f}^2
\label{comp} \label{freeze}
\eeq
The freeze-out condition implies the scaling $T_f^3 \sim \sqrt{N}$. 
From Eqs.~(\ref{nperp}) and (\ref{freeze}), it is then clear that changes
in
$N$, caused  for example by a change in the number of light neutrinos
$N_\nu$,  would directly influence $n/p$, and hence the $^4$He
abdundance.  The dependence of the light element abundances on $N_\nu$ is
shown in Figure~\ref{fig:3} \cite{cfo2}, where plotted is the mass
fraction of 
$^4$He, $Y$, and the abundances by number of the D, $^3$He, and $^7$Li 
as a function of the baryon-to-photon ratio, $\eta$,  for values of
$N_\nu = 2 - 7$. As one can see, an upper limit to $Y$, combined with a
lower limit to $\eta$ will yield an upper limit to $N_\nu$ \cite{ssg}. 

Assuming no new physics at low energies, the value of $\eta$ is the 
sole input parameter to BBN calculations. It is fixed by the comparison 
between BBN predictions and the observational determinations of the 
isotopic abundances~\cite{cfo1}. From $^4$He and $^7$Li, one finds a 
relatively low value~\cite{Fields:1996yw,cfo1} for 
$\eta \sim 2.4 \times 10^{-10}$ 
corresponding to a low baryon density $\Omega_B h^2 = 0.009$ with a 
95\% CL range of 0.006 -- 0.017.  Deuterium, on the other hand, implies 
a large value of $\eta$ and hence a large baryon density: 
$\eta \sim 5.8 \times 10^{-10}$ 
and $\Omega_B h^2 \sim 0.021$ with a 95\% CL range of 0.018 -- 0.027.
The value of the baryon density has also been determined recently from
measurements of microwave background anisotropies. The recent result
from DASI~\cite{dasi} indicates that $\Omega_B h^2 = 
0.022^{+0.004}_{-0.003}$, while that of  BOOMERanG-98~\cite{newboom}, 
$\Omega_B h^2 = 0.021^{+0.004}_{-0.003}$ (using 1$\sigma$ errors).

\begin{figure}
\begin{center}
\includegraphics[height=6.0cm]{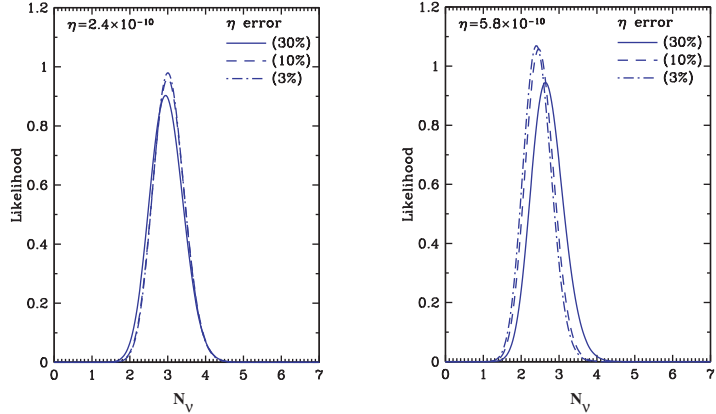}
\vskip-0.5truecm
\caption{(a)
The distribution in
$N_\nu$ assuming a value of $\eta = 2.4 \times 10^{-10}$ from
$^4$He and $^7$Li and the CBI measurement of the microwave background
anisotropy \protect\cite{cfo2}. The curves show the effect of the expected
increased accuracy in the CMB determination of $\eta$. (b) 
(b)  As in (a), but assuming a value of $\eta = 5.8 \times 10^{-10}$ from
D and the DASI and BOOMERanG measurements of the microwave background
anisotropy.}
\label{fig:4}      
\end{center}
\end{figure}

With the value of $\eta$ fixed, one can use He abundance measurements 
to set limits on new physics. In particular one can set upper limits
the number of neutrino flavors. Taking $Y_p = 0.238 \pm 0.002 \pm
0.005$ (see e.g. \cite{yp}), we show in Figure~\ref{fig:4} the likelihood
functions for $N_\nu$ based on both the low and high values of $\eta$
\cite{cfo2}. The curves show the impact of an increasingly accurate
determination of $\eta$ from 30\% to 3\%. If one assumes a 20\%
uncertainty in $\eta$ (the current uncertainty level), these 
calculations provide upper limits of
\begin{eqnarray}
N_\nu < 3.9 & \qquad \eta = 2.4 \times 10^{-10} \nonumber \\
N_\nu < 3.6 & \qquad \eta = 5.8 \times 10^{-10} 
\label{BBNnubounds}
\end{eqnarray}
at the 95\% CL. Although, as noted above, LEP has already placed very 
stringent limit to $N_\nu$, the limit~(\ref{BBNnubounds}) is useful,
because it actually applies to the total number of new particle degrees
of freedom and is not tied specifically to neutrinos.  In fact, more
generally, the neutrino limit can be translated into a limit on the
expansion rate of the Universe at the time of BBN, which can be applied 
to a host of other constraints on particle properties.

\subsection{BBN limits on Neutrino masses and lifetimes}
\label{sec:1.3.1}

As discussed above, the nucleosynthesis prediction for light element 
abundances is sensitive to the changes in the expansion rate of the 
universe, which depends on the energy density of the universe during 
the BBN era (\ref{H}). This extra energy density could be in the form 
of new massless degrees of freedom, in which case their number is 
directly constrained by equation (\ref{BBNnubounds}). Equally well 
the extra energy density could reside in the form of massive long 
lived but unstable neutrinos, in which case nucleosynthesis provides
interesting constraints on their masses and life-times.

We already pointed out that the relic density of neutrinos strongly 
depends on whether they decouple while relativistic or nonrelativistic. 
Here the calculations also depend on how the neutrino life-times relate
to the BBN time-scale of about 100 seconds.  Also, in order to get 
reliable results for the light element abundances, one must keep track
of the induced perturbations (electron neutrino heating) in the weak 
reaction rates (\ref{weakrates}) in addition to computing changes in 
the expansion rate. Nevertheless, even in this case it is customary to 
measure the change in helium abundance in units of equivalent effective 
neutrino degrees of freedom $N_{\rm eff}$, such that the limit 
(\ref{BBNnubounds}) can be applied on 
$N_{\rm eff}(\Delta Y(m_\nu,\tau_\nu))$.

\begin{figure}
\begin{center}
\includegraphics[height=5.5cm]{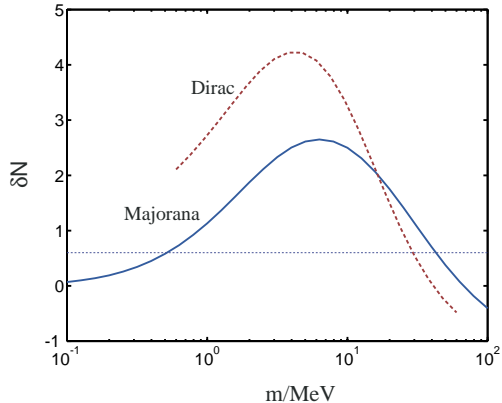}
\caption{Plot of the effective number of neutrino degrees of freedom
         during BBN for a Dirac (dashed) and a Majorana neutrino 
         (solid)~\protect\cite{FKOandKK}.}
\label{fig:5}       
\end{center}
\end{figure}

When the neutrino life-time is much larger than 100 seconds, neutrinos
are effectively stable on a nucleosynthesis 
scale~\cite{Kolb:1991sn,FKOandKK,HaMaDoHa}. 
While accounting for changes 
in the rates in Eq. (\ref{weakrates}) is important for the detailed
bounds,  the bulk behaviour of $N_{\rm eff}(m_\nu)$ is dictated by the
neutrino mass contribution to the energy density. Obviously, when $m_\nu
\ll 0.1$ MeV,  neutrinos are effectively massless during BBN, and $N_{\rm
eff}(m_\nu) 
\rightarrow 3$ when $m_\nu \rightarrow 0$.  For masses in excess of 
$0.1$ MeV, but below the neutrino decoupling temperature of 
${\cal O}({\rm few})$ MeV, their number density is unsuppressed 
and their mass density can be large during BBN, causing $N_{\rm eff}$ 
to increase. For yet larger masses however, the Boltzmann factor shown 
in (\ref{annrate}) begins to suppress the mass density
and eventually turns $N_{\rm eff}$ down again. This behaviour is 
shown in Figure~\ref{fig:5} for a massive Dirac and Majorana type
tau-neutrino~\cite{FKOandKK}.
The bound (\ref{BBNnubounds}) yields an excluded region for
stable neutrino masses centered around ${\cal O}({\rm few})$ MeV. 
For $N_\nu < 3.6$ the lower bound is $m_\nu > 42$ MeV (Majorana)
and $m_\nu > 30$ MeV (Dirac)~\cite{FKOandKK}. This is only 
relevant for $\nu_\tau$, and is complementary to the present 
laboratory limit on the $\tau$-like neutrino, $m_{\nu} < 18$
MeV~\cite{taulab}. Due to  contributions from pion decays and inverse
decays to neutrinos, the  upper bound from BBN depends on the QCD-phase
transition temperature, 
$T_{\rm QCD}$, and is also different for $\tau$- and $\mu$-neutrinos 
because of their different scattering rates off muons. 
Imposing again the constraint $N_\nu < 3.6$, and taking 
$T_{QCD}= 200~{\rm MeV}$ gives~\cite{DKR}:
\begin{eqnarray}
m_{\nu}  & \la &  230~{\rm KeV} \qquad {\rm \mu-like} \nonumber \\ 
m_{\nu} & \la &  290~{\rm KeV} \qquad {\rm \tau-like}.
\end{eqnarray}
The laboratory limit on the muon-like neutrino, for comparision, is
$m_{\nu} \la 170$ keV~\cite{mubound}. To improve this, the BBN
limit should be improved to $N_\nu \la 3.4$~\cite{DKR}.

\begin{figure}
\begin{center}
\includegraphics[height=7cm]{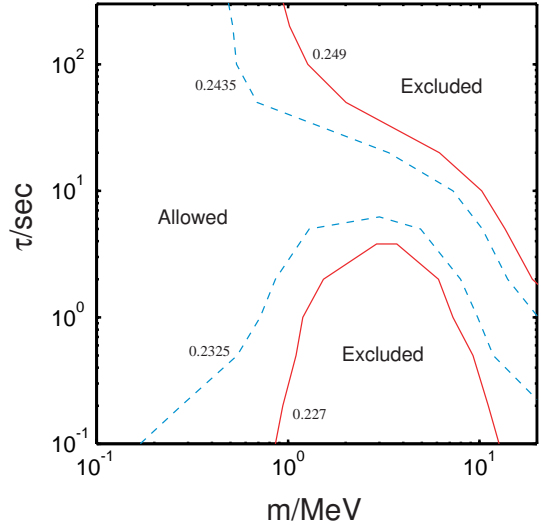}
\caption{Plot of BBN-constraint on masses and life-times of an unstable
tau neutrino.  Contours are labeled by one (dashed) and two sigma 
(solid) deviations from the observed value of $Y_p$. The upper 
right corner is excluded due to too much and the lower part of the 
graph by too little $^4$He being produced.}
\label{fig:6}      
\end{center}
\end{figure}

When the neutrino life-time is small or comparable to the nucleosynthesis
time scale, one has to account for neutrino decay processes as well.
This  involves solving for the distributions of the final state decay
products, which might include new particles like majorons, and their
possible direct  effect on BBN. (For example energetic photons would
cause the dissociation of the newly generated light nuclei.)  Such
calculations have been done by  many groups~\cite{groupref,DolHan}, and
the results are given in exclusion  plots in the mass-vs-life-time plane.
Constraints are possible for masses of order 
${\cal O}(1)$ MeV and life-times of order ${\cal O}(1)$ second. 
In Figure~\ref{fig:6}, we show a constraint on 
tau-neutrino masses and life-times as an example (data taken from the 
reference~\cite{DolHan}).

\subsection{BBN limits on Neutrino mixing parameters}
\label{1.3.2}

Despite losing the competitive edge {\em w.r.t.~} masses and 
life-times, BBN continues to put interesting limits on other neutrino 
mass parameters, relevant for neutrino oscillations. Indeed, the 
LEP-limit of course only applies to neutrinos with weak interactions, 
while neutrinos without weak interactions, or {\em sterile} neutrinos, 
have been proposed in many different contexts over the years. At 
present, a prime motivation to introduce sterile neutrinos is to 
explain the LSND neutrino anomaly~\cite{LSND} in conjunction 
with the solar and atmospheric neutrino deficits.

\begin{figure}
\begin{center}
\vskip 0.5truecm
\includegraphics[height=6.5cm]{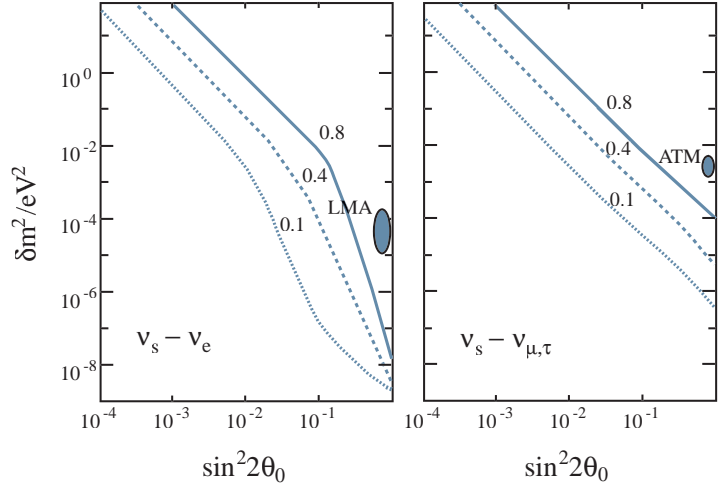}
\caption{Plotted are the BBN constraints on the active sterile 
neutrino mixing parameters~\protect\cite{EKTBig} 
for $\nu_\alpha = \nu_e$ (left) and
$\nu_\alpha = \nu_{\mu,\tau}$ (right). Regions to the right from
the contours labeled by the bound on effective neutrino degrees 
of freedom $\delta N_{\rm eff} = N_{\rm eff}-3$ are excluded.
Shown are also the current regions corresponding to active-sterile
mixing parameters for atmospheric (ATM) and the large mixing angle 
(LMA) solar neutrino solutions.}
\label{fig:7}
\end{center}
\end{figure}

BBN on the other hand is sensitive to any type of energy density 
changing the expansion rate in the ${\cal O}$(0.1-1) MeV range,
irrespective of their interactions. It is therefore very interesting 
to observe that even if no sterile neutrinos were created at very early 
times, they could be excited by mixing effects in the early universe 
prior to nucleosynthesis. The basic mechanism is very simple. Suppose
that an active state $\nu_{\alpha}$ ($\alpha = e,\mu,\tau$) {\em mixes} 
with a sterile state $\nu_s$. That is, the neutrino mass matrix and 
hence the Hamiltonian, is not diagonal in the interaction bases. The
mixing  is further affected by the forward scattering interactions with
the  background plasma, felt by the active state.  As a result, even
though  a neutrino state was initially produced in purely active
projection,  after some time $t$ it has become some coherent linear
combination of  both active and sterile states:
\begin{equation}
  \nu(t) = c_{\rm e}(t)\nu_{\rm e} + c_{\rm s}(t)\nu_{\rm s}.
\label{mixture}
\end{equation}
The coherent evolution of this state is interrupted by collisions,
which effect a sequence of quantum mechanical measurements of the 
flavour content of the propagating state. Since the sterile state has 
no interactions, each measurement is complete, and collapses the 
wave-function to the sterile state with a probability 
$P_{\nu_{\rm e} \rightarrow \nu_{\rm s}}(t) = {| c_{\rm s}(t) |}^2$.  
As a result, the sterile states are populated roughly with an 
average rate
\begin{equation}
\Gamma_{\nu_{\rm s}} = \Gamma_{\nu}{\langle 
{| c_{\rm s}(t) |}^2\rangle}_{\rm coll} = 
\frac{1}{2}\sin^22\theta_m \Gamma_{\nu_\alpha}. 
\end{equation}
where $\Gamma_{\nu_\alpha}$ is the weak interaction rate of the 
active state $\nu_\alpha$, and we have assumed that the oscillation 
time is short in comparision with the collision time scale. The 
matter mixing angle $\theta_m$ is given by~\cite{EKTBig}
\begin{equation}
\sin^22\theta_{\rm m} = \frac{\sin^22\theta_0}
                       {{1-2 \chi\cos2\theta_0 + \chi^2}}
\end{equation}
where $\sin2\theta_0$ is the mixing angle in vacuum and 
$\chi\equiv 2p|V|/\delta m^2$ where $\delta m^2$ is the mass 
squared difference of the vacuum mass eigenstates, $p\sim T$ is 
the  momentum and $|V|$ is the matter induced effective potential 
to the Hamiltonian~\cite{Raffelt,EKTBig}. The weak rate scales as 
$\Gamma_{\nu_\alpha} \sim T^5$. Moreover $\chi \sim T^6$ at very 
high temperatures, which causes a strong matter suppression for 
mixing and hence $\Gamma_{\nu_{\rm s}} \sim T^{-7}$. At very small 
temperatures $\theta_m \rightarrow \theta_0$ on the other hand, and 
hence $\Gamma_{\nu_s} \sim T^5$. The rate is thus suppressed both at 
very large and at very small temperatures~\cite{Kainulainen:bn}. 
In the intermediate region of ${\cal O}({\rm few})$ MeV however,
$\Gamma_{\nu_{\rm s}}$ can exceed the expansion rate bringing a
significant amount of sterile neutrinos into equilibrium. An 
accurate treatment of the problem requires a numerical solution 
of the appropriate quantum kinetic equations, and the results 
depend on whether the mostly active state is heavier ($\delta 
m^2 <0$) or lighter ($\delta m^2 >0$) of the mixing states. 
We show the results of such a calculation in Figure~\ref{fig:7} 
below~\cite{EKTBig}. The lines are labeled by constant effective 
number of degrees of freedom during BBN: $\delta N_\nu \equiv 
N_\nu - 3$. The most recent limits corresponding to (\ref{BBNnubounds})
can be interpolated from the curves shown.
It is the area above  the curves which is excluded by BBN-limit.

The BBN-limit can be converted to an upper bound on the sterile 
neutrino flux~\cite{Kainulainen:2001cb} in the atmospheric and the 
solar neutrino observations. Using $N_\nu < 3.6$ one finds
\begin{eqnarray}
\sin^2\theta_{\mu \rm s} &\la 0.03& \qquad \rm (Atmospheric) \nonumber\\ 
\sin^2\theta_{\rm es}    &\la 0.06& \qquad \rm (Solar \;\; LMA), 
\label{fluxbounds}
\end{eqnarray}
whereas the bounds from the atmospheric and solar neutrino experiments
are about an order of mangitude weaker: $\sin^2\theta_{\mu \rm s} \la 
0.48$ and $\sin^2\theta_{\rm es} \la 0.72$.

The constraints shown in the Figure~\ref{fig:7} and in Equation
(\ref{fluxbounds}) depend on the assumption that the primordial 
lepton asymmetry is not anomalously large~\cite{Enqvist:1990dq}. 
In ref.~\cite{australia} it was suggested that a large effective  
asymmetry violating this assumption could actually be generated by 
oscillations given a particular neutrino mass and mixing hierarchy. 
For a while these ideas generated a lot of interest, as they would 
have allowed reconciling all observed anomalies (including LSND) 
with the nucleosynthesis constraints. 
However, in these scenarios at least one of the active states would
have to be much heavier than the two others, which is not allowed by 
the atmospheric and solar neutrino flux observations. As a result, 
the bounds (\ref{fluxbounds}) hold, and in particular nucleosynthesis 
is very much at odds with the possible existence of the LSND-type 
sterile state. To see this, observe that creating a large enough 
effective mixing between $\bar \nu_\mu$ and $\bar \nu_{\rm e}$ to 
explain the anomaly~\cite{LSND} would require a sterile intermediare 
with $m_{\nu_s} \simeq 1$ eV, and $\sin^22\theta_{\mu\rm e} \simeq 
\frac{1}{2}\sin^22\theta_{\mu\rm s} \sin^22\theta_{\rm se} \ga 
10^{-2}(\delta m^2/{\rm eV}^2)^{-2}$. In other words at least 
one of  the active-sterile mixings should satisfy 
$\sin^22\theta_{(\mu, {\rm e})\rm s} 
\ga 0.15(\delta m^2/{\rm eV}^2)^{-1}$, which is well within the 
BBN excluded regions shown in Figure~\ref{fig:7}. 
(It would be excluded even by $N_{\rm eff} \la 3.9$, although
we do not show that contour in Figure~\ref{fig:7}).

It should finally be noted that active-active type oscillations 
have hardly any effect on the expansion rate or the weak interaction 
rates~\cite{Langacker:1986jv}, and hence are not constrained in the
above sense by BBN. However, large mixing angle active-active 
oscillations could equilibrate the lepton {\em asymmetries} prior 
to BBN. This has been used to put strong bounds on muon- and 
tau-lepton asymmetries \cite{Lbounds}, which exclude the possibility 
of the degenerate nucleosynthesis.

\section{Neutrinos and Supernovae}
\label{sec:1.4}

Neutrinos have for long been known to play important role in the
physics of supernovae.  To be sure, it is clear that by far the largest 
part, roughly 99$\%$, of the gravitational binding energy of about 
$3\times 10^{53}$ erg involved in the explosion of a type II supernova 
is carried out by neutrinos, while just 1$\%$ powers the shock wave 
responsible for blowing out the mantle of the star, and only a tiny 
fraction of about 0.1$\%$ escapes in the form of light responsible 
for the spectacular sights observed in the telescopes watching the sky.

\begin{figure}
\begin{center}
\includegraphics[height=6cm]{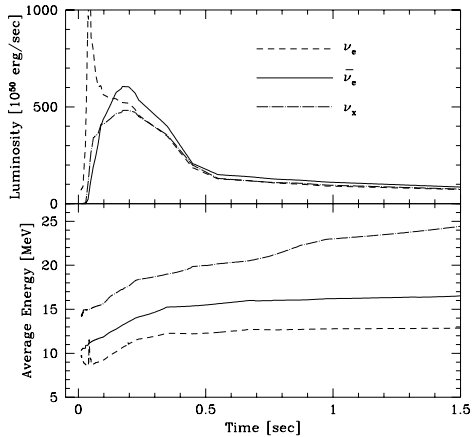}
\caption{Time evolution of neutrino luminosities and average energies:
$\nu_x$ represents the spectrum of $\nu_\mu$, $\nu_\tau$, 
$\bar \nu_\mu$ and $\bar \nu_\tau$. Figure taken
from~\cite{futurenuexps}. }
\label{fig:8}      
\end{center}
\end{figure}

The formation of the neutrino burst in the collapse of a type II 
supernova is rather well understood. The temporal structure of the 
burst and the energy spectrum of the emitted neutrinos can be computed
fairly well~\cite{Giorg}. Existing or planned large scale neutrino 
detectors~\cite{futurenuexps} can be used to observe deviations from 
these signatures and to obtain interesting information on neutrino
masses and mixing parameters~\cite{SmirnovandX}, given a future 
observation of a Galactic supernova.  We show an example of a 
compilation of neutrino fluxes and spectra in Figure~\ref{fig:8}.

A number of constraints on new physics and in particular on neutrino
parameters have already been deduced from the famous SN1987A event in 
the Small Magellanic Cloud. Of these, perhaps the most direct is 
the upper bound on the neutrino mass 
derivable from the maximum duration of the observed duration of the
neutrino pulse of about 10 seconds. Given the initial energy 
spectrum of neutrinos and the distance to the supernova, one can 
compute the expected spread in the arrival times of the neutrinos to 
earth as a function of neutrino mass. Comparing the predictions with
the observations has been shown to yield the bound~\cite{SNnumassbnd}
\begin{equation}
  m_{\nu_e} \la 6 - 20 \; {\rm eV}.
\end{equation}

The observed pulse length also lends 
to the classic {\em cooling argument}: any new physics that would 
enhance the neutrino diffusion such that the cooling time drops below 
the observed duration, must be excluded.  Cooling arguments have been 
used to set limits on various neutrino properties~\cite{giorgbook}
such as active-sterile neutrino mixing~\cite{itejamuut} and neutrino 
magnetic moments. The magnetic dipole moment bound in particular was 
recently revised by Ayala etal.~\cite{Ayala} to
\begin{equation}
\mu_{\nu_e} \la 1-4 \times 10^{-12} \mu_B
\label{mubound}
\end{equation}
which is a couple of orders of magnitude more stringent than the
best laboratory bounds, and comparable to the bound coming from the 
globular cluster red giant cooling  arguments~\cite{Raffelt:gv}, 
which give $\mu_{\nu_e} \la 3 \times 10^{-12} \mu_B$.

At present, most of the activity concerning neutrinos in supernovae 
has focussed on the effects of neutrino transport 
in supernova explosion dynamics rather than with finding constraints
on neutrino mixing  parameters or masses. Indeed, the details 
of the physics responsible for the actual visibly observed supernova 
explosion, including blowing out the stellar mantle, are not very 
well understood. In particular, the shock wave, which forms deep 
within the iron core as the infall of matter is reversed due to the
stiffening of the nuclear matter equation of state, is typically 
found to be too weak to explode the star. This is believed to be due to
energy loss from the shock and dissociating iron
nuclei on its way out from  the core to the mantle. In the popular
``delayed explosion scenario", the stalling shock wave is rejuvenated by
energy transfer to the  shock from the huge energy-flux of neutrinos
free streaming away from  the core. Recent numerical simulations
including diffusive neutrino  transport do not verify this expectation
however~\cite{bosnbang};  while neutrinos definitely help, they do not
appear to solve the  problem. These results are not conclusive
because the  diffusive transport equations used so far~\cite{bosnbang}
did not  include all relevant neutrino interactions, most notably the
nuclear brehmsstralung processes~\cite{nubrehm,Giorg}. Furthermore, 
processes other than diffusive processes, such as convective 
flows in the core and behind the shock, appear to play an important 
role as well~\cite{giorgreview}.

It is of course possible that a succesfull SN explosion requires help 
of some new physics to channel energy more efficiently to the shock,
and neutrino-oscillations have already been considered for this 
role~\cite{fulleretal}. The idea is that $\nu_\mu$ and $\nu_\tau$,
interact more weakly and hence escape more energetic from deeper
in the core than do electron neutrinos.
%(see Figure~\ref{fig:8}). 
Arranging mixing parameters so that $\nu_{\mu,\tau}$ turn resonantly 
to $\nu_{\rm e}$ in the mantle behind the shock could increase the 
energy deposited to the shock significantly. Unfortunately the mass 
difference needed for the resonant transition would be very big:
\begin{equation}
\delta m^2 \sim 2 E V_{\rm eff} \simeq 1.5 \times 10^5 
\rho_e E_\nu^{100}\; {\rm eV}^2
\end{equation}
where $\rho_e$ is the electron density in units 
$10^{10}\,\rm g/cm^3$, which at the shock front is about $10^{-3}$,
and $ E_\nu^{100}$ is the neutrino energy in units of 100 MeV.
So one gets $m_\nu \ga 10$ eV, which is excluded by the present 
data.  A similar idea was behind the suggestion \cite{Voloshin88}, that
neutrino magnetic moments induce resonant transitions from
$\nu_R$ (which escape energetic from deep in the core) to $\nu_L$ behind
the mantle. This mechanism  could actually be used to evade the bound
(\ref{mubound}), but as it  demands a magnetic moment of order $\mu_\nu
\sim 10^{-11}\mu_B$ it  has problems in coping with the red giant cooling 
bound~\cite{Raffelt:gv}.

Sterile neutrinos could also be relevant for supernovae by alleviating
the problems with the r-process nucleosynthesis, which is thought 
to be responsible for creation of the most heavy elements. In the 
standard SN calculations the r-process nucleosynthesis is not 
effective due to too efficient de-neutronization by the processes
$\nu_e + n \rightarrow e^- + p$. If electron neutrinos mixed with a 
sterile state however, these processes could be made less effective, 
increasing the neutron density in the mantle, and hence improving 
the r-process efficiency~\cite{rprocSN}.

Finally, there is also the old problem of the ``kick"-velocities of 
pulsars (neutron star remnants of supernova explosions). It has proven 
difficult to arrange for these velocities, which average around
450 km/sec, based on the normal fluid dynamics in asymmetrical 
SN-explosions. The momentum carried out 
by neutrinos $p_\nu \simeq E_\nu$, on the other hand, is about 100 
times larger than the pulsar kinetic energy, so that a mere one per 
cent asymmetry in the neutrino emission would be enough to power 
the pulsar velocities. Interesting attempts have been made to 
explain such asymmetric emission by an asymmetric distribution of 
inhomogeneities in the SN magnetic fields, combined with a large
neutrino magnetic moment~\cite{Voloshin88}, or just the magnetic 
field induced deformation of the neutrino spheres~\cite{kusenkosegre}. 
However, the former would again need probably too large magnetic moment 
to work, and a detailed analysis of the latter suggests that the a
symmetric flux is very suppressed, requiring perhpas unrealistically 
large magnetic fields; according to~\cite{jankaraffelt} the field
needs to be in excess of $10^{17} G$, while \cite{KusenkoLater} argue
that a field of  $10^{14-15} G$ would suffice. The true nature of the
physics explaining the kick velocities may remain ambiguous for some
time to come, but a neutrino solution looks definitely appealing from 
the pure energetics point of view. 

Before we conclude this section, we would like to mention, several 
other astrophysical limits on neutrino properties. A sure limit to the 
mean life/mass ratio is obtained from solar x- and $\gamma$- ray 
fluxes~\cite{raffmean}. The limit is $\tau/m_{\nu_1} > 7
\times 10^9$ s/eV.  This is far superior to the laboratory bound of 300
s/eV~\cite{reines}. Other much stronger limits ($> O(10^{15}$) s/eV) are
available from the lack of observation of $\gamma$-rays in coincidence
with neutrinos from SN 1987A~\cite{other}. This latter limit applies to
the heavier neutrino mass eigenstates, $\nu_2$ and $\nu_3$, as well.

\section{Neutrinos and Cosmic Rays}
\label{sec:1.5}

\begin{figure}
\begin{center}
\includegraphics[height=6.5cm]{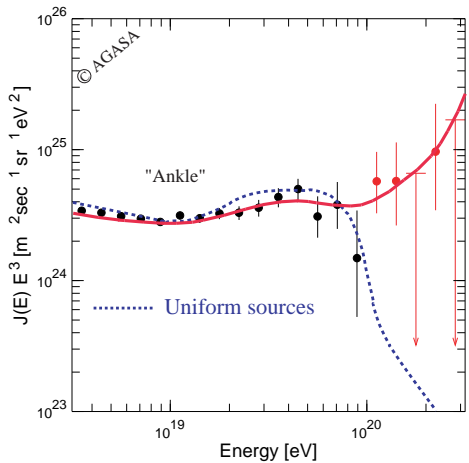}
\vskip-0.4truecm
\caption{Sceled by $E^3$ spectrum of highest energy cosmic rays near
the GZK cutoff. The dotted line corresponds to the expectation from
uniformly distributed extragalactic sources, and the solid line shows
the prediction of a $Z$-burst model of ref.~\cite{gelmini}. 
(Figure modified from the orginal found at the AGASA web 
page~\cite{Agasa}.)}
\label{fig:9}      
\end{center}
\end{figure}

One of the most interesting puzzles in astrophysics today concerns the
observations of ultra high energy cosmic rays (UHECR), beyond the so 
called Greisen-Zatsepin-Kuzmin (GZK) cutoff
\begin{equation}
  E_{\rm GZK} \simeq 5 \times 10^{19} \, \rm eV.
\end{equation}
The problem is that cosmic rays at these energies necessarily need
to be of extragalactic origin, since their gyromagnetic radius within the 
galactic magnetic field far exceeds galactic dimensions. Yet,
the attenuation lengths of both protons and photons are rather small
in comparision with intergalactic distances, and neither can have 
originated by further than about $50$ Mpc away from us, due to their
scattering off the intergalactic cosmic photon background. As a result, 
one would expect that the cosmic ray spectrum would abruptly end 
around $E \sim E_{\rm GZK}$ due to scattering off of the microwave
background. This cutoff is represented by the dotted line in 
Figure~\ref{fig:9}. In contrast, several groups, most notably the 
HiRes~\cite{hires} and AGASA~\cite{Agasa} collaborations, have  reported
events with energies well above the GZK-cutoff: the latest  compliation
of AGASA, for example, contains 10 events above the scale $E > 10^{20}$
eV, observed since 1993.

The origin of UHECR's has been the subject of lively 
discussions over the last few years. Apart from having astrophysical 
orgins, being acclerated in extreme environments at extragalactic 
distances (in AGN's, GRB's or Blazars), they have been attributed 
to the decay products of very heavy particles or of topological 
defects. All these explanations have problems, however.
Astrophysical explanations face the difficult task of accelerating
particles to the extreme energies required, with little or no
associated sub-TeV-scale photonic component (as none have been
ever observed). This is in addition to the 
above mentioned problem of the propagation of UHECR's over 
extragalactic distances.
Decay explanations are somewhat disfavoured by the growing 
evidence of doublets and triplets in AGASA data, and by the 
correlation between the UHECR arrival directions with far compact 
Blazars~\cite{tinyakov}, which appear rather to be pointing towards 
astrophysical origin.

The attenuation problem for extragalactical UHECR's can be avoided 
however, if they cross the universe in the form of a neutrino beam, 
since neutrinos travel practically free over super-Hubble distances.
Indeed, should this be the case, the initial $\nu_{\rm UHE}$'s could
occasionally interact with the cosmological relic neutrino background 
close to us, giving rise to ``$Z$-bursts" of hadrons and
photons~\cite{weiler}, which then would give rise to the observed 
UHECR events. Indeed, given a neutrino mass $m_\nu$,
such a collision has sufficient CM-energy for resonant $Z$ production 
if
\begin{equation}
E_\nu = \frac{M^2_Z}{2m_\nu} \simeq 4.2 \times 10^{21} {\rm eV}
  \left(  \frac{{\rm eV}}{m_\nu} \right).
\end{equation}
The requirement of super-GZK-energies for the
initial $\nu_{\rm UHE}$ beam leads immediately to the interesting 
mass scale for neutrinos: $m_\nu \sim {\cal O}(1)$ eV.

$Z$-burst models have been extensively studied lately~\cite{fodor}.
For example, in ref.~\cite{gelmini} it was shown that a $Z$-burst 
model with $m_\nu = 0.07$ eV, corresponding to a degenerate neutrino 
spectrum, could  reproduce the AGASA-data including the spectral features,
such as  the ``ankle" and the ``bump" observed at $E \la E_{\rm GZK}$. 
The best fit model of ref.~\cite{gelmini} is shown by the solid line
in Figure~\ref{fig:9}. 
While the agreement with AGASA data is good, it should be noted that
this model predicts that cosmic ray primaries are exclusively photons 
above $E \ga 10^{20}$ eV, whereas the $E \simeq 3 \times 10^{20}$ eV 
event observed by Fly's Eye is almost certainly not caused by a 
photon~\cite{Haltzen}. The model also predicts a large increase of the 
cosmic ray flux above few$\times 10^{20}$ eV (as 
a direct result of the huge initial energy needed: $E_{\nu_{\rm UHE}}
\simeq 6 \times 10^{22}$ eV),  which excacerbates the already outstanding
problem of the origin  of UHECR's. These problems could be ameliorated by
assuming somewhat larger neutrino masses, and it has been argued by Fodor 
{\em etal.~}\cite{fodor}, that the $Z$-burst scenario can already
be used to {\em constrain} the mass, plausibly to within $m_\nu 
\sim 0.08 - 1.3$ eV, in very good agreement with other mass 
determinations. The analysis of ref.~\cite{fodor}, was restricted 
to $Z$-resonance interactions however, while for higher CM-energies 
the cross section~\cite{Enqvist:1988we} for pair production of gauge 
bosons $\nu\nu\rightarrow ZZ,WW$ becomes important. These reactions
have been shown to give accptable solutions with larger neutrino 
masses $m_\nu \ga 3$ eV~\cite{fargion}.

In summary, although the origin of UHECR's remains a mystery today, 
it is almost certain that if it has to do with extragalactic sources, 
neutrino physics plays a very important role in the interpretation of
these events.

\section*{Acknowledgments}

K.K.\ would like to thank Steen Hannestad and Petteri Keranen for 
useful advice. The work of  K.A.O. was supported partly by DOE grant 
DE--FG02--94ER--40823.

\newpage
%
%\vspace*{3.0cm}

%
\end{document}